\documentclass[aip, pof,graphicx]{revtex4-2}

\usepackage{graphicx}
\usepackage{dcolumn}
\usepackage{bm}
\usepackage[utf8]{inputenc}
\usepackage[T1]{fontenc}
\usepackage{mathptmx}
\usepackage{color}
\usepackage{longtable}
\usepackage{changes}

\draft 

\begin{document}


\title[Growth of barchan dunes of bidispersed granular mixtures]{Growth of barchan dunes of bidispersed granular mixtures\\
This article may be downloaded for personal use only. Any other use requires prior permission of the author and AIP Publishing. This article appeared in Physics of Fluids 33, 051705 (2021) and may be found at https://doi.org/10.1063/5.0048696} 



\author{Carlos A. Alvarez}
\altaffiliation[Also at ]{Department of Atmospheric and Oceanic Sciences, University of California, Los Angeles,\\
	Los Angeles, CA 90095-1565, USA}
\author{Fernando David C\'u\~nez}
\author{Erick M. Franklin}%
 \email{erick.franklin@unicamp.br}
 \thanks{Corresponding author}
\affiliation{ 
School of Mechanical Engineering, UNICAMP - University of Campinas,\\
Rua Mendeleyev, 200, Campinas, SP, Brazil
}%


\date{\today}

\begin{abstract}
Barchans are dunes of crescentic shape found on Earth, Mars and other celestial bodies, growing usually on polydisperse granular beds. In this Letter, we investigate experimentally the growth of subaqueous barchans consisting of bidisperse grains. We found that the grain distribution within the dune changes with the employed pair, and that a transient stripe appears on the dune surface. We propose that observed patterns result from the competition between fluid entrainment and easiness of rolling for each grain type, and that grains segregate with a diffusion-like mechanism. Our results provide new insights into barchan structures found in other environments.
\end{abstract}


\maketitle 


Barchan dunes, or simply barchans, are crescent-shaped dunes with horns pointing downstream that are frequently found on Earth, Mars  and other celestial bodies \cite{Bagnold_1, Herrmann_Sauermann, Hersen_3, Elbelrhiti, Claudin_Andreotti, Parteli2, Courrech}. They are formed under one-directional flows and limited amount of available grains and, under these conditions, present a robust shape that arises in diverse environments with a large range of scales, going from the orders of the kilometer and millennium in the case of Martian dunes down to centimeters and minutes in the aquatic case \cite{Hersen_1, Claudin_Andreotti, Yang_3}.

Because of their ubiquitous nature, barchans have been studied from field measurements on Earth \cite{Finkel, Long, Norris2, Hesp, Sauermann_1, Yang_2} and Mars \cite{Breed, Schatz, Bishop, Bourke, Chojnacki, Runyon}, which usually involve large time and length scales and uncontrolled conditions. Those studies produced valuable data that increased our understanding of many aspects of barchan morphodynamics, though with limited applicability for bedform evolution over long times and little information at the grain scale. Given the smaller and faster scales of subaqueous barchans, experimental investigations were carried out in water channels and tanks under controlled conditions, from which it was possible to obtain the initial and long-time evolutions of the barchan morphology \cite{Hersen_1, Alvarez} and the motion of individual grains on the barchan surface \cite{Alvarez3, Alvarez4, Wenzel}. However, although natural beds are composed of polydisperse grains \cite{Bagnold_1, Finkel, Long}, the previous experiments investigated barchans of monodisperse particles, the only bidisperse-related references being the experiments of Caps and Vandewalle \cite{Caps} and Rousseaux et al. \cite{Rousseaux} for size segregation within two-dimensional ripples and those of Groh et al. \cite{Groh2} for density segregation within two-dimensional dunes.

Variations in the morphology and structure of barchans have been observed in nature, being described as resulting from hilly terrains \cite{Finkel, Bourke, Parteli4}, wind changes \cite{Finkel, Bourke, Parteli4}, and barchan collisions \cite{Long, Hersen_5, Bourke, Vermeesch, Parteli4, Assis}. The effect of polydispersity (even bidispersity) on the barchan structure is less understood. On Earth, the previous studies on the sedimentary structure of barchan dunes were conducted with field measurements, and focused on the impressions left by the long-time migration of aeolian barchans \cite{Kocurek2, Bristow_charlie, Gomez-Ortiz, Fu, Bristow_Charles}. Those studies, by interpreting the strata, provide an account of the past history of barchan fields; however, because those structures result from long-time processes that involve erosion and accretion of sand, wind variations, and, in some cases, water content, it is difficult to isolate and understand the real effect of polydisperse grains in stratification. On Mars, dunes with regions of different color, sometimes over the dune surface and other times on peripheral regions, were observed on its north pole and have been associated with the presence of coarse-grained ice together with sandy sediments \cite{Breed, Schatz, Feldman, Ewing, Hansen, Hansen2, Portyankina, Pommerol}. In these cases, the presence of two species with different sizes and densities could induce concentration of one of the species around the barchans, or induce segregation in horizontal layers. Still, the precise organization of grains in those two-species barchans remains unknown.

In this Letter, we investigate experimentally the growth of subaqueous barchans consisting of bidisperse grains. We carried out exhaustive experiments where two-species barchans, in terms of grain sizes and/or densities, were tracked, and we varied the grain properties (size and density), their concentrations, and the water flow rate. We found that, while the barchan morphology remains roughly constant, denser, smaller, and smaller and less dense grains tend to accumulate on the barchan surface. For higher concentrations of the species accumulating on the surface, the other one forms a bottom layer and appears in top-view images as being around the barchan. We also found the appearance of a transient line on the dune surface, transverse to the flow direction, that forms during the growth of the single barchan, upstream its crest, and migrates toward the leading edge until disappearing. When grains of different densities but same size are used, that line consists in a large transverse stripe with a high concentration of the lighter grains. For the other pairs, that line consists in a transverse line that clearly separates a downstream region where segregation is complete from an upstream one where segregation is ongoing. The appearance of the transient line is intriguing, and we propose that it is the result of a competition between the strength of fluid entrainment and easiness of rolling for each grain type. For grains of different sizes, we observe, when the transient is finished, the presence of oblique stripes of much smaller wavelength than the dune size, which we associate with size segregation.

\begin{figure}
    \begin{center}
     \includegraphics[width=.95\linewidth]{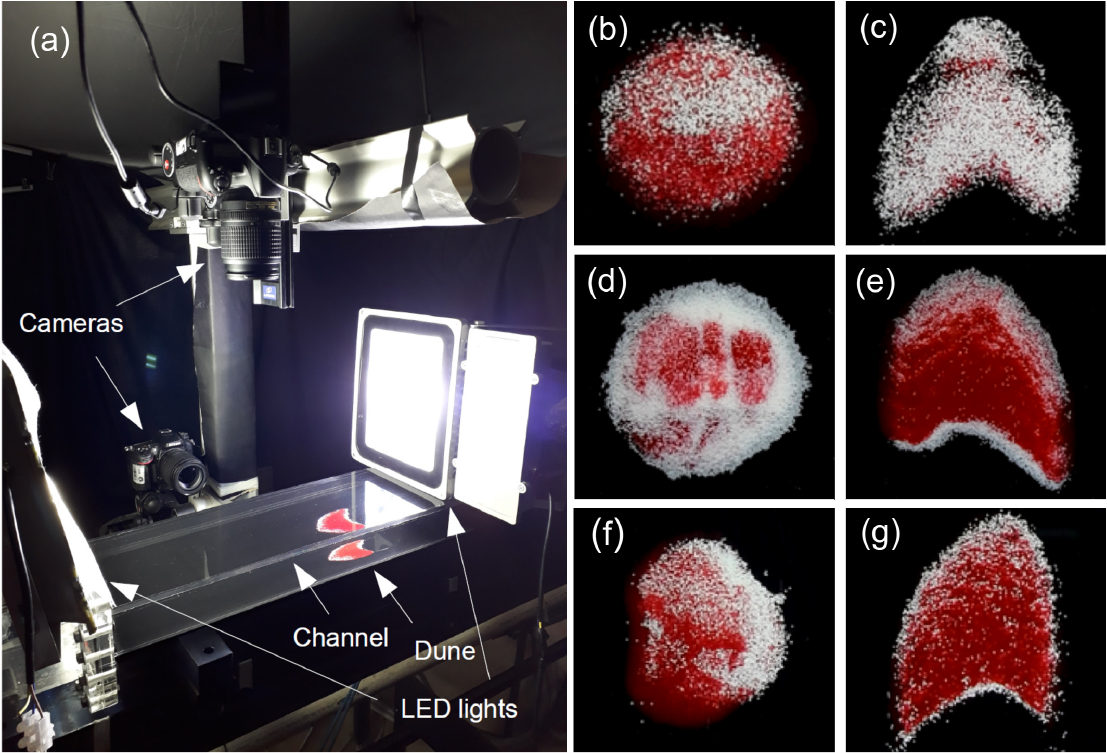}\\
    \end{center}
\caption{(a) Photograph of the experimental setup showing the test section, cameras, LED lights and dune. (b) and (c) Top views of the initial pile and barchan dune, respectively, for case e of Table \ref{tab1} (different densities and same diameter); (d) and (e) initial pile and barchan dune, respectively, for case h of Table \ref{tab1} (same density and different diameters); (f) and (g) initial pile and barchan dune, respectively, for case q of Table \ref{tab1} (different densities and diameters). In Figures (b) to (g), flow is from top to bottom (Multimedia view).}
\label{fig:photos}
\end{figure}

The experimental device consisted of a water reservoir, two centrifugal pumps, a flow straightener, a 5-m-long closed-conduit channel of transparent material and rectangular cross section (width = 160 mm and height 2$\delta$ = 50 mm), a settling tank, and a return line. The last 2 m of the channel consisted of the 1-m-long test section followed by a 1-m-long section discharging in the settling tank. With the channel previously filled with water, mixtures of grains were poured inside, forming an initial pile of conical shape that was deformed afterward into a barchan dune by imposing a turbulent water flow. It is important to note that during the pouring of grains some segregation took place, so that each initial condition was unique in terms of granular structure within the pile. We used different grain sizes and densities that were mixed by pairs in different concentrations, and we varied the water flow rate. By repeating some test conditions, we obtained barchans that showed the same patterns (described in the following), indicating that variations in local concentrations in the initial pile do not affect the resulting barchan. A camera placed above the channel acquired images of the bedforms, from which measurements at both the dune and grain scales were obtained by image processing \cite{Alvarez, Alvarez3, Alvarez4, Alvarez_these, Kelley, Crocker, Cunez3}, while a camera placed horizontally acquired side-view images used for measurements of the dune height. With this technique, we had access to the granular structure over the barchan surface, but not to its inner structure. Figure \ref{fig:photos}a shows a picture of the test section, and the layout of the experimental device and details about the used cameras are available in the supplementary material.

\begin{table}[!ht]	
	\begin{center}
		\begin{tabular}{c c c c c c c c c c}
			\hline\hline
			Case & $\phi_1$ & $\phi_2$ & $\phi_3$ & red & Re & $d_{ratio}$ & $\rho_{s,ratio}$ & $m_0$ \\ 
			$\cdots$  & $\cdots$ & $\cdots$ & $\cdots$ & $\cdots$ & $\cdots$ & $\cdots$ & $\cdots$ & (g) \\\hline
			a & 0.8 & 0.2 & 0 & $S_2$ & 1.47 $\times$ 10$^4$ & 1 & 1.64 & 15.5 \\
			b & 0.5 & 0.5 & 0 & $S_2$ & 1.47 $\times$ 10$^4$ & 1 & 1.64 & 13.7 \\
			c & 0.2 & 0.8 & 0 & $S_2$ & 1.47 $\times$ 10$^4$ & 1 & 1.64 & 11.8 \\
			d & 0.8 & 0.2 & 0 & $S_2$ & 1.82 $\times$ 10$^4$ & 1 & 1.64 & 15.5 \\
			e & 0.5 & 0.5 & 0 & $S_2$ & 1.82 $\times$ 10$^4$ & 1 & 1.64 & 13.7 \\
			f & 0.2 & 0.8 & 0 & $S_2$ & 1.82 $\times$ 10$^4$ & 1 & 1.64 & 11.8 \\
			g & 0 & 0.8 & 0.2 & $S_3$ & 1.47 $\times$ 10$^4$ & 2.5 & 1 & 10.5 \\
			h & 0 & 0.5 & 0.5 & $S_3$ & 1.47 $\times$ 10$^4$ & 2.5 & 1 & 10.5 \\
			i & 0 & 0.2 & 0.8 & $S_3$ & 1.47 $\times$ 10$^4$ & 2.5 & 1 & 10.5 \\
			j & 0 & 0.8 & 0.2 & $S_3$ & 1.82 $\times$ 10$^4$ & 2.5 & 1 & 10.5 \\
			k & 0 & 0.5 & 0.5 & $S_3$ & 1.82 $\times$ 10$^4$ & 2.5 & 1 & 10.5 \\
			l & 0 & 0.2 & 0.8 & $S_3$ & 1.82 $\times$ 10$^4$ & 2.5 & 1 & 10.5 \\
			m & 0.8 & 0 & 0.2 & $S_3$ & 1.47 $\times$ 10$^4$ & 2.5 & 1.64 & 15.5 \\
			n & 0.5 & 0 & 0.5 & $S_3$ & 1.47 $\times$ 10$^4$ & 2.5 & 1.64 & 13.7 \\
			o & 0.2 & 0 & 0.8 & $S_3$ & 1.47 $\times$ 10$^4$ & 2.5 & 1.64 & 11.8 \\
			p & 0.8 & 0 & 0.2 & $S_3$ & 1.82 $\times$ 10$^4$ & 2.5 & 1.64 & 15.5 \\
			q & 0.5 & 0 & 0.5 & $S_3$ & 1.82 $\times$ 10$^4$ & 2.5 & 1.64 & 13.7 \\
			r & 0.2 & 0 & 0.8 & $S_3$ & 1.82 $\times$ 10$^4$ & 2.5 & 1.64 & 11.8 \\\hline
		\end{tabular}
	\end{center}
	\caption{Label of tested cases, initial concentration (volume basis) of each species within the initial pile, species of red (darker) color, channel Reynolds number $Re$, ratio between grain diameters $d_{ratio}$, ratio between grain densities $\rho_{s,ratio}$, and mass of the initial heap $m_0$.}
	\label{tab1}
\end{table}

The tests were performed with tap water at temperatures between 21 and 25 $^o$C and three populations of particles were used: round zirconium beads (grain density $\rho_s = 4100$ kg/m$^3$) with grain diameters within $0.40$ mm $\leq\,d\,\leq$ $0.60$ mm, round glass beads ($\rho_s$ = 2500 kg/m$^3$) with 0.40 mm $\leq\,d\,\leq$ 0.60 mm, and round glass beads with 0.15 mm $\leq\,d\,\leq$ 0.25 mm, which we call species 1, 2 and 3 ($S_1$, $S_2$ and $S_3$), respectively, and for which $\phi_1$, $\phi_2$ and $\phi_3$ are the used concentrations (see the supplementary material for microscopy images of the used grains). For each dune, grains of different species had either white or red colors in order to track their distribution over the barchan dune. The cross-sectional mean velocities of water $U$ were 0.294 and 0.364 m/s, corresponding to Reynolds numbers based on the channel height, Re = $\rho U 2\delta /\mu$, of $1.47 \times 10^4 $ and $1.82 \times 10^4 $, respectively, where $\rho$ is the density and $\mu$ the dynamic viscosity of the fluid. The shear velocities on the channel walls in the absence of dunes, $u_*$, were computed from velocity profiles measured with a two-dimensional two-component particle image velocimetry device (2D2C-PIV), and were found to follow the Blasius correlation \cite{Schlichting_1}. By using the hydraulic diameter of the channel, $u_*$ is found to be 0.0168 and 0.0202 m/s for the two flow rates employed, corresponding to variations of Reynolds numbers at the grain scale, Re$_*$ = $\rho u_* d / \mu$, within 3 and 10 and to Shields numbers, $\theta$ = $(\rho u_*^2)/((\rho_s - \rho )gd)$, within 0.02 and 0.14, where we considered the midrange mean of the diameter of each species and $g$ is the acceleration of gravity. The values of $u_*$ correspond to a base state on the channel wall, the shear velocity over a moving barchan reaching a maximum value of approximately 1.4$u_*$ close to its crest \cite{Andreotti_2, Franklin_11}. All initial heaps had a volume of 7.0 cm$^3$, with initial masses varying between 10.5 and 15.5 g. The tested conditions are summarized in Table \ref{tab1}, where $d_{ratio}$ and $\rho_{s,ratio}$ are the ratios between grain diameters and densities, respectively (see the supplementary material for values of Re$_*$ and $\theta$ and pictures of initial and final bedforms for each test). 

\begin{figure}[ht]
    \begin{center}
     \includegraphics[width=.99\linewidth]{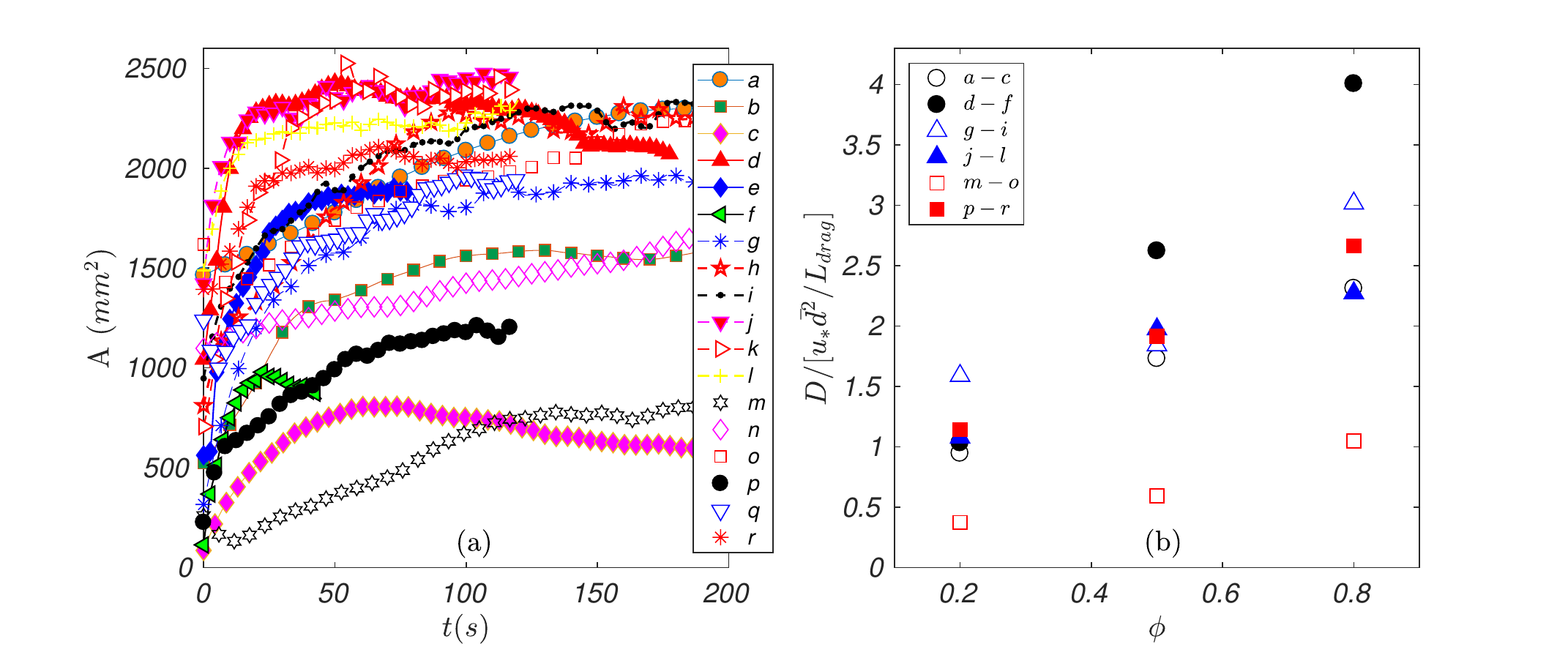}\\
    \end{center}
\caption{(a) Projected area $A$ of the dune occupied by the species accumulating over its surface as a function of time. Cases c and m had their time scale multiplied by 0.65 and 0.35, respectively, in order to fit the graphic. (b) Normalized diffusion-like coefficient $D$ as a function of the concentration $\phi$ of the species accumulating over the surface. Cases are listed in the key.}
	\label{fig:diffusion_areas}
\end{figure}

The general aspects of our results are summarized in Figs. \ref{fig:photos}b to \ref{fig:photos}g (Multimedia view). For different mixtures, grains segregate forming two main layers, one over the other, where the top layer consists of (i) denser grains ($S_1$) when different densities are used ($S_1$ and $S_2$, Figs. \ref{fig:photos}b and \ref{fig:photos}c); (ii) smaller grains ($S_3$) in the case of different diameters ($S_2$ and $S_3$, Figs. \ref{fig:photos}d and \ref{fig:photos}e); and (iii) smaller and less dense grains ($S_3$) for different diameters and densities ($S_1$ and $S_3$, Figs. \ref{fig:photos}f and \ref{fig:photos}g). Those results are surprising, different gradings having been reported in the literature. For different sizes, cases g to r of Table \ref{tab1}, measurements on two-dimensional subaqueous ripples \cite{Caps, Rousseaux} showed that larger grains tend to accumulate over their crest, while coarse sediments form a bottom layer that remains immobile in the case of river dunes \cite{Kleinhans}. In the present case, although forming a bottom layer, coarse grains are periodically entrained by the fluid flow. For grains of different densities and same size, cases a to f of Table \ref{tab1}, a startling result had already been found by Groh et al. \cite{Groh2} for two-dimensional dunes, where denser particles were found to accumulate in the central body of the dune, close to its top but covered by a thin layer of less dense grains. An explanation for that grading remains missing, and now we find something still different in the case of barchans. In the present case, we propose, based on Makse \cite{Makse}, that the observed grading is the result of a competition between grains more easily entrained by the fluid flow and those that roll more easily over the bed, as explained in the following. Interestingly, the segregation in horizontal layers could be at the origin, together with solidification and frosting (as proposed by Refs. \cite{Hansen, Hansen2, Portyankina, Pommerol}), of part of the structures observed on barchans on the north pole of Mars, for which top-view images show that dunes consisting of coarse-grained ice together with sandy sediments present one species sometimes over the dune surface and other times on peripheral regions.

For different densities ($S_1$, $S_2$), the less dense grains ($S_2$) are entrained more easily by the fluid and, therefore, the Shields number is the pertinent parameter: the species with higher $\theta$ presents higher mobility ($S_2$), migrating toward the lee face and forming afterward a carpet for the other species ($S_1$), so that they ($S_2$) accumulate at the bottom. For different diameters ($S_2$, $S_3$), the smaller grains ($S_3$) have higher $\theta$, being entrained more easily by the fluid, but they have also smaller Reynolds numbers. In our experiments, both the ratios of $\theta$ and $Re_*$ between the two species were approximately 2.5. Because $Re_*$ is equivalent to a dimensionless roughness \cite{Schlichting_1}, a layer consisting of the smaller species ($S_3$) implies higher mobility of the other species ($S_2$) when compared with the inverse. Therefore, if we consider a mixed bed, smaller grains tend to be shielded from the flow even if they present higher values of $\theta$, while larger ones roll easily over smaller grains. This balance was analyzed by Yager et al. \cite{Yager}, who showed that protrusions of certain species within a given polydisperse bed reduce significantly the critical Shields number for that bed (as a whole). Because in the bidisperse case the protrusions are related with $Re_*$ and the flow entrainment with $\theta$, an analysis based on $Re_*$ and $\theta$ seems adequate. Within the range of parameters of our experiments, the results show that the effect caused by the grain size is stronger and $Re_*$ is the pertinent parameter: the species with higher $Re_*$ ($S_2$) presents higher mobility and accumulates at the bottom. In our experiments with different sizes and densities ($S_1$, $S_3$), smaller and less dense grains ($S_3$) presented values of $Re_*$ 2.5 times smaller and $\theta$ 5 times greater than those of denser and larger grains ($S_1$). Although the ratio for $\theta$ is two times stronger than that for $Re_*$, the easiness of rolling produces a stronger effect and the species with higher $Re_*$ ($S_1$) forms the bottom layer. In summary, our results show that the bottom layer is composed of grains with higher $Re_*$ and, in case of equal $Re_*$, with higher $\theta$.

\begin{figure}[ht]
    \begin{center}
     \includegraphics[width=.9\linewidth]{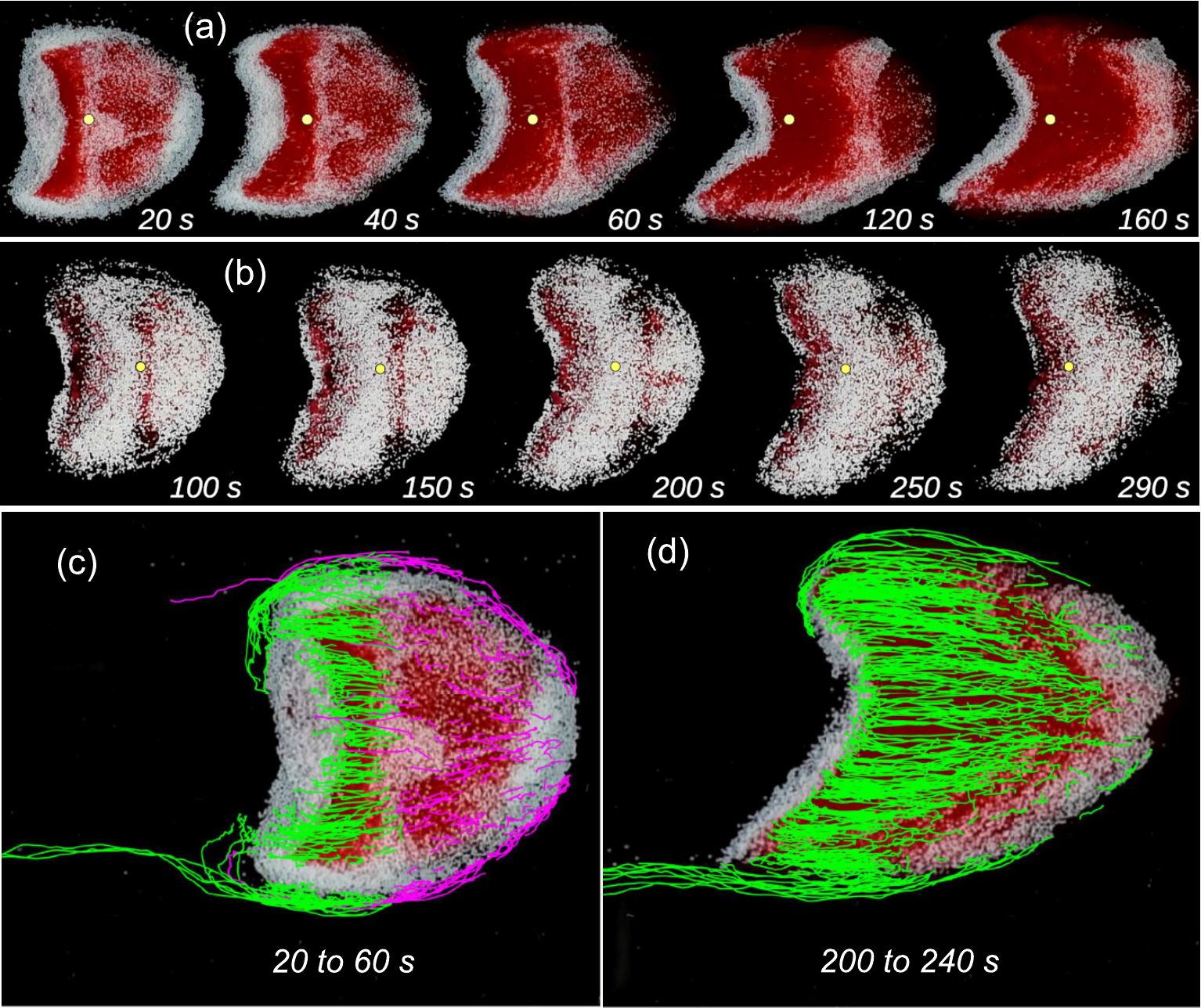}\\
    \end{center}
	\caption{Evolution of the transverse stripe and trajectories of grains. (a) and (b) Snapshots of barchans at different instants showing the transverse stripe as it migrates to the leading edge. They correspond to cases h and b of Table \ref{tab1}, respectively, and the yellow circle indicates the crest position. In Figure (a), times correspond to 1.0, 2.0, 3.0, 6.0 and 8.0$t_c$, and in Figure (b) to 1.2, 1.8, 2.4, 3.0 and 3.5$t_c$. (c) and (d) Trajectories of larger grains of case h, for two different intervals: when the stripe was still present and when it had already vanished, respectively. Trajectories were computed over 40 s and are superposed with the bedform image at the initial time of their computation (from 1 to 3$t_c$ in Figure (c) and from 10 to 12$t_c$ in Figure (d)). In Figure (c), green lines correspond to grains that started moving in positions downstream the transverse stripe, while magenta lines correspond to those that started moving in positions upstream the stripe.}
	\label{fig:transverse_stripe}
\end{figure}

In order to evaluate the segregation in horizontal layers, we analyzed the increase in the area occupied by grains accumulating over the dune surface. Figure \ref{fig:diffusion_areas}a presents the projected area $A$ of the dune (top view) occupied by the species accumulating over its surface as a function of time, for all the tested conditions, and it is a measure of the spreading of those species over the surface. Figure \ref{fig:diffusion_areas}a shows the existence of an initial phase consisting of a positive slope, in which the spreading of grains over the surface is occurring at a roughly constant rate, and a final phase consisting of a plateau, reached when the quantity of grains of each species over the surface remains roughly constant. The initial phase has a duration equal or greater than the characteristic time for the displacement of barchans, $t_c$, computed as the dune length divided by its celerity \cite{Alvarez}. This means that a fast spreading takes place during one or more turnovers of the bedform, when all the grains get mobilized at least once, and afterward the segregation persists. A table showing the duration of the initial phase of each test normalized by $t_c$ is available in the supplementary material. We note that the same kind of migration toward the dune surface was not observed for growing barchans consisting of a single species \cite{Alvarez3, Alvarez4, Alvarez5}.

Although grains are entrained by the fluid, we associate the spreading of grains over the surface with a diffusion-like mechanism, in a similar way as in the turbulence viscosity \cite{Schlichting_1}. The accumulation of grains on certain regions of a granular bed, with its consequent deformation, has been described recently as a diffusion process by Abramian et al. \cite{Abramian} in the case of monodisperse grains. We make use of a similar idea, but we evaluate diffusion in terms of the general spreading of one species over the other. For that, we defined a diffusion coefficient $D$ (mm$^2$/s), given by the positive slope of the initial phases of each curve in Fig. \ref{fig:diffusion_areas}a, and which we expect to scale, as shown in Eq. \ref{Eq:normal_D}, with the inertial length for the stabilization of sand flux \cite{Hersen_1} $L_{drag}$ = $( \bar{\rho} / \rho ) \bar{d}$, the shear velocity $u_*$, and the mean diameter weighted by concentrations $\bar{d} = \left( \frac{\phi_{a}}{d_a} + \frac{\phi_{b}}{d_b} \right) ^{-1}$, where $\bar{\rho} = \phi_a \rho_a + \phi_b \rho_b$, and $a$ and $b$ correspond to each species. Figure \ref{fig:diffusion_areas}b presents $D \left( u_* \bar{d}^2 / L_{drag} \right) ^{-1}$ as a function of particle concentration for all tested cases, for which we obtain diffusion-like coefficients of the same order of magnitude, around unit. Although there is some variation, a diffusion-like coefficient seems proper for predicting the transient duration for the spreading of a species over the dune surface.

\begin{equation}
D \sim \frac{u_* \overline{d}^2}{L_{drag}}
\label{Eq:normal_D}
\end{equation}

The general structures observed over barchans after segregation has taken place can be better understood from the transverse line that forms on the dune surface during the growth of the single barchan and migrates toward the leading edge until disappearing, while slowly migrating downstream with respect to the channel wall (their celerities with respect to the channel wall varied between 0.07 and 1.43 mm/s, while those of the dune centroid were between 0.19 and 2.18 mm/s). The evolution of the transverse stripe is shown in Fig. \ref{fig:transverse_stripe} together with trajectories of larger grains migrating over the dune. Despite showing only some trajectories in Figs. \ref{fig:transverse_stripe}c and \ref{fig:transverse_stripe}d, they represent well the behavior of the whole set. During the transient, we can observe a region where segregation has already occurred, downstream the transverse line, where larger (higher $Re_*$) and lighter (higher $\theta$) grains, in cases of either bidispersed size or density, respectively, roll/jump over longer distances. These grains arrive directly on the crest, from where they fall by avalanches over the lee face, and horns, while the same does not occur on the region upstream the transverse line. In the latter region, segregation occurs faster close to the transverse line, where larger (higher $Re_*$ in cases g to r) or lighter (higher $\theta$ in cases a to f) grains hop to the already segregated surface, and from there roll/jump easier over the other species (see the supplementary material for average lengths and velocities of rolling/jumping grains upstream and downstream the transverse stripe). With that process, while grains with higher $Re_*$ or $\theta$ (if equal $Re_*$) travel to the lee face and form a carpet for the other ones, in a mechanism similar to the one proposed by Groh et al. \cite{Groh2} in the 2D case, the line is displaced toward the leading edge until segregation is complete. When grains of different densities but same size are used ($S_1$, $S_2$), that line consists of a large transverse stripe with a high concentration of lighter grains ($S_2$, that appears as a dark stripe in Fig. \ref{fig:transverse_stripe}b). The origin of the transverse stripe is always close to the crest of the initial pile, downstream of which segregation occurs faster due to the formation of a recirculation region and lee face, and from that position it propagates upstream with respect to the dune. The formation of a transverse stripe due to grain bidispersity in a barchan dune can be at the origin of some stratified structures found in nature. For example, a previous flooding could bring different species to a barchan dune, that would then develop a transverse stripe. Therefore, considerations about bidispersity should be taken into account when describing the history of barchan fields based on the analysis of strata.

Finally, we observed also the presence of oblique stripes with wavelengths scaling with the grain diameter, as can be observed in the upstream region of the dune shown in Fig. \ref{fig:photos}e. Based on images of cases g, h, j and k, we computed the arithmetic mean of wavelengths of oblique stripes, $\lambda$, 20 s after the transverse stripe had vanished, and found values of the order of 10$d$ (12 $\leq$ $\lambda /d$ $\leq$ 16, see the supplementary material for more images showing the oblique stripes and a table with $\lambda$ for each case measured). It is difficult to explain their formation without measuring the inner grading of barchans; however, we speculate that their origin is a size segregation similar to those found in avalanches \cite{Makse2}, but with the driving force in the direction of water streamlines. Here again, strata found in nature may reflect the oblique stripes caused by size segregation, which must thus be taken into account in order to avoid misinterpretation.

In conclusion, we observed a general segregation with one particular species covering most of the dune surface, and that, although grains are entrained by the fluid, the spreading of that species over the surface can be analyzed as a diffusion-like mechanism. We showed that the amount of grains accumulating over the surface grows initially at a constant rate and eventually reaches a plateau, and proposed a diffusion coefficient, $D \sim u_* \overline{d}^2 {L_{drag}^{-1}}$, for the initial growth. We observed also a transient line on the dune surface, transverse to the flow direction, that migrates toward the leading edge, and that, when grains of different densities but same size are used, consists of a large transverse stripe. We propose that both the transverse line and the segregation in horizontal layers result from a competition between the strength of fluid entrainment and easiness of rolling for the involved species. In general, grains with higher $Re_*$ form the bottom layer of bidispersed barchans, but when the different species have close values of $Re_*$ that layer consists of grains with higher $\theta$. While segregation is still incomplete, the transverse stripe can be observed. Finally, we noticed the formation of oblique stripes for grains of different sizes, which we speculate to be associated with size segregation. Our findings open new possibilities to explain barchan structures and stratification found in other environments. For example, part of the stratification of barchans observed on the north pole of Mars could result from size and density segregation. Another example is the use of strata to deduce the history of barchan fields on Earth, which must consider the influence of grain dispersity in the formation of different layers.

\section*{SUPPLEMENTARY MATERIAL}
See the supplementary material for the layout of the experimental device, microscopy images of the used grains, tables listing dimensionless groups, diffusion times, wavelengths of oblique stripes and average lengths and velocities of rolling/jumping grains, snapshots of barchans at initial and final stages for different grain types, and trajectories of grains over the barchan dune.

\section*{DATA AVAILABILITY}
The data that support the findings of this study are openly available in Mendeley Data on https://data.mendeley.com/datasets/z42c97sw4c, Ref. \cite{Supplemental3}.

\begin{acknowledgments}
C. A. Alvarez is grateful to SENESCYT (Grant No. 2013-AR2Q2850) and to CNPq (Grant No. 140773/2016-9), F. D. C\'u\~nez is grateful to FAPESP (Grant No. 2016/18189-0), and E. M. Franklin is grateful to FAPESP (Grant No. 2018/14981-7) and to CNPq (Grant No. 400284/2016-2) for the financial support provided.
\end{acknowledgments}

\bibliography{references}

\end{document}